\documentclass[]{jfm}
\usepackage{float}
\usepackage{graphicx}
\usepackage{epstopdf,epsfig}
\usepackage[colorlinks=true, citecolor=blue, linkcolor=blue, urlcolor=red]{hyperref}
\usepackage{amsmath}
\usepackage{amssymb}
\usepackage{wasysym}
\usepackage{tikz}
\usepackage{subfigure}
\usepackage{ragged2e}

\newcommand{\tikzcircle}[2][red,fill=red]{\tikz[baseline=-0.75ex]\draw[#1,radius=#2] (0,0) circle ;}%

\usepackage{color}
\definecolor{color1}{rgb}{0.3686,0.3098,0.6235}
\definecolor{color2}{rgb}{0.3892,0.6422,0.6676}
\definecolor{color3}{rgb}{0.7484,0.8971,0.7121}
\definecolor{color4}{rgb}{0.9065,0.3601,0.2686}
\definecolor{color5}{rgb}{0.6157,0.0039,0.2588}

\shorttitle{Bubbly flows at small Re}
\shortauthor{M. Ravisankar and R. Zenit}

\title{Velocity fluctuations for bubbly flows at small Re}

\author{Mithun Ravisankar
 \and Roberto Zenit
 }

\affiliation{School of Engineering, Brown University, 184 Hope St, Providence, RI 02912, USA
}

\begin{document}

\maketitle

\begin{abstract}

We experimentally investigate the effect of Reynolds number (Re) on the turbulence induced by the motion of bubbles in a quiescent Newtonian fluid at small Re. The energy spectra, $E(k)$, is determined from the decaying turbulence behind the bubble swarm obtained using particle image velocimetry (PIV). We show that when Re $\sim$ $O$(100), the slope of the normalized energy spectra is no longer independent on the gas volume fraction and the $k^{-3}$ subrange is significantly narrower, where $k$ is the wavenumber. This is further corroborated using \textcolor{black}{second-order longitudinal velocity structure function and }spatial correlation of the velocity vector behind the bubble swarm. On further decreasing the bubble Reynolds number ($O(1) < $ Re $ < O(10)$), the signature $k^{-3}$ of the energy spectra for the bubble induced turbulence is replaced by $k^{-5/3}$ scaling.  Thus, we provide experimental evidence to the claim by \citet{mazzitelli2003effect} that at low Reynolds numbers the normalized energy spectra of the bubble induced turbulence will no longer show the $k^{-3}$ scaling because of the absence of bubble wake and that the energy spectra will depend on the number of bubbles, thus non-universal.

\end{abstract}


\section{Introduction}\label{sec:intro}

When a swarm of bubbles rises in an otherwise stagnant fluid due to the motion of bubbles disturbances are created in the surrounding fluid giving rise to velocity fluctuations. At a moderate to high Reynolds number, Re, these velocity fluctuations in the wake behind the bubbles interacts with one another giving rise to the emergence of $k^{-3}$ scaling of the energy spectra, $E(k)$, of velocity fluctuations. This signature $k^{-3}$ scaling of the energy spectra instead of the classical $k^{-5/3}$ Kolmogorov's scaling is regarded as the bubble induced turbulence, also referred to as pseudoturbulence \citep{risso2018agitation}. 

\citet{lance1991turbulence} were the first to observe the emergence of $k^{-3}$ scaling of energy spectra, $E(k)$, with respect to the wavenumber, $k$ for the bubble induced turbulence in Newtonian fluids. They argued that in a spectral space, the balance between the energy produced by the motion of bubbles and the viscous dissipation in a statistically steady state gives rise to the emergence of $k^{-3}$ scaling. Following this pioneering work, a number of numerical \citep{esmaeeli1996inverse,bunner2002dynamics2,balachandar2010turbulent} and experimental \citep{zenit2001measurements,martinez2007measurement,riboux2013model,prakash2016energy,almeras2017experimental} works supported the emergence of $k^{-3}$ scaling for both the spatial and temporal velocity fluctuations generated by the bubble motion \citep{risso2018agitation}. \citet{amoura2017velocity} showed experimentally that irrespective of the dispersed phase, be it bubbles or random fixed solid spheres, the velocity fluctuations generated in the continuous phase gives rise to the $k^{-3}$ scaling. \citet{pandey2023kolmogorov} identified the coexistence of Kolmogorov's turbulence with the bubble induced turbulence for a wide range of Reynolds number and Galilei number (ratio of buoyancy to viscous forces). Recently, \citet{zamansky2024turbulence} proposed that the $k^{-3}$ subrange of the energy spectra results from the mean shear rate imposed by the bubbles, ruling out the speculation by \citep{lance1991turbulence} that the balance between the spectral production and the spectral dissipation.

Experimentally determining the liquid velocity fluctuations in two-phase gas-liquid flows is challenging. Due to the dispersed nature of the bubbly flow, laser-based techniques within the bubble swarm can only be used for very dilute flows. On the other hand, the hot-wire-based techniques \citep{martinez2007measurement,mendez2013power,almeras2017experimental}, are far from perfect as their use implies elaborate signal processing. To overcome these difficulties, \citet{riboux2010experimental} measured the liquid velocity fluctuations by abruptly stopping the bubble formation using a solenoid valve thus leaving the wake behind the bubble swarm free of bubbles to be analyzed using particle image velocimetry (PIV). They showed that the $k^{-3}$ scaling , independent of the gas volume fraction and bubble diameter, is observed between the Eulerian length scale ($\Lambda = D/C_d$) and the bubble diameter ($D$), where $C_d$ is the drag coefficient of a single rising bubble. Later \citet{risso2018agitation} modified the Eulerian length scale, $\Lambda = {D}/{\sqrt{C_d\text{Re}}}$ to include the Reynolds number. 

An intriguing question is to explore what happens to the energy spectra of the bubbly flows at lower Reynolds number \citep{bunner2002dynamics1,bunner2002dynamics2}. When Re $<$ 20, the bubbles will not have a significant wakes behind them \citep{blanco1995structure,mougin2001path}. A clue on what to expect for smaller Reynolds numbers was summarized by \citet{mazzitelli2003effect}, who conducted numerical simulations for bubbly Newtonian fluids considering bubbles as point particles. Specifically, they did not observe the $k^{-3}$ scaling. Thus concluding the essentiality of the wakes. They further argued that the energy spectrum slope for bubbly flows with little to no wakes behind them will depend on the number of bubbles and will therefore be non-universal \citep{mazzitelli2003effect,mazzitelli2009evolution}. Most surprisingly, to our knowledge, there are only a few experimental studies in this regime \citep{cartellier2001bubble,martinez2007measurement,mendez2013power}. \citet{cartellier2001bubble} studied bubble induced agitations at Re $\sim$ $O$(1); however, to delay the onset of large scale instabilities (transition into heterogeneous bubbly regime) an inner tube within the bubble column was used to create the liquid flow rate due to the gas lift. Further, they did not address the nature of energy spectra of liquid velocity fluctuations. 

In the present study we will focus only on the homogeneous bubbly regime to address the effect of Reynolds number on the bubble induced turbulence. Using the particle image velocimetry technique proposed by \citet{riboux2010experimental}, the liquid velocity fluctuations are determined to calculate the energy spectra. We find that as the Reynolds number decreases the slope of the energy spectra emerges as $k^{-5/3}$ instead of the $k^{-3}$ scaling observed for the bubble induced turbulence. Further, we show that this slope of the energy spectra depends on the number of bubbles in the column. Understanding the spectral structure fluid velocity fluctuations in two-phase flows at low Re could be justified by their relevance in modern applications such as microfluidics \citep{anna2016} and hydrogen production \citep{avci2022}.

\section{Experimental setup and methods}\label{sec:setup}

\subsection{Experimental setup}

\begin{figure}[h]
\centering
       \includegraphics[scale=0.45]{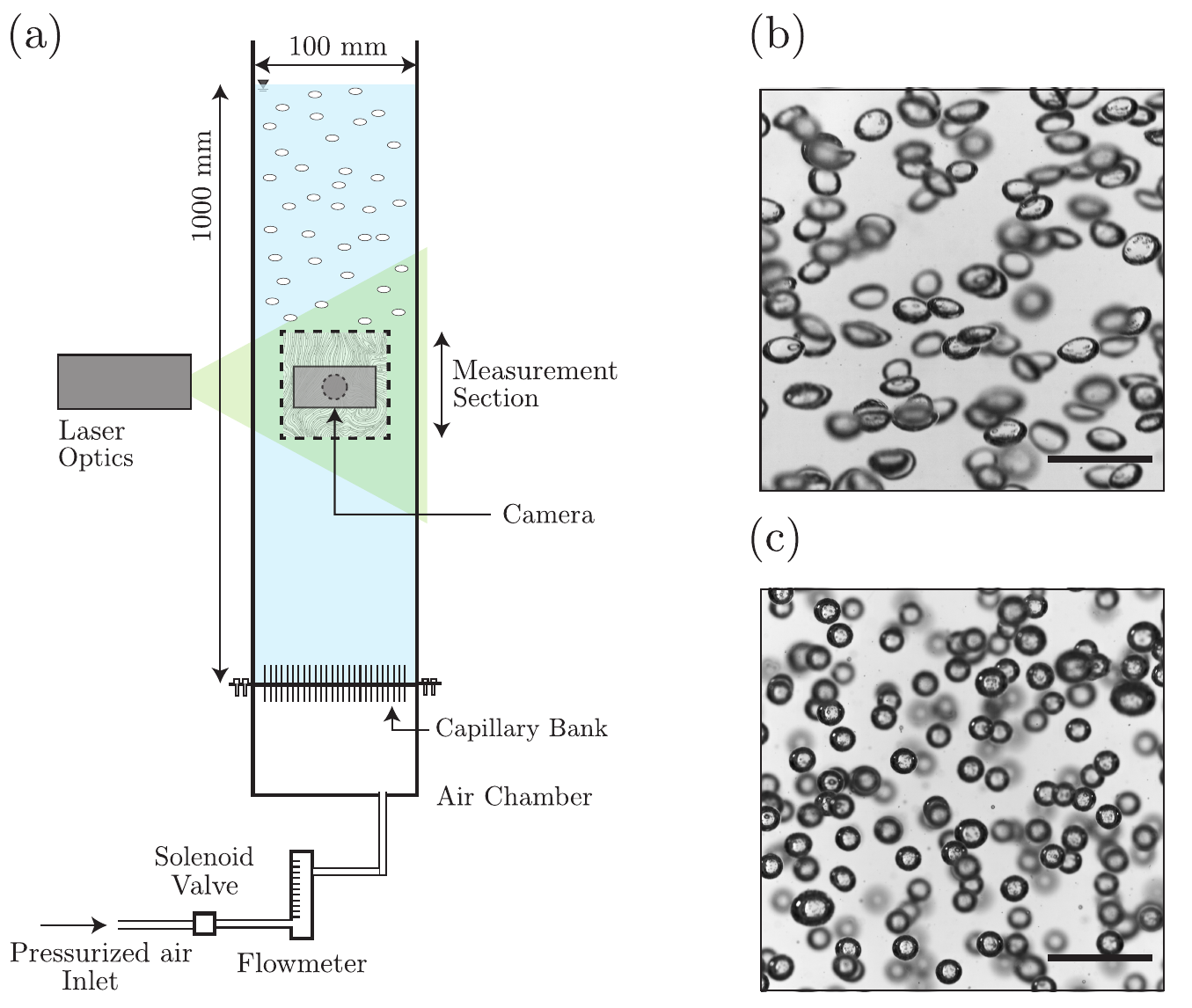}
    \caption{\justifying (a) Schematic of the experimental setup used in the current study. Using a solenoid valve, the formation of bubbles are abruptly stopped and the wake behind the bubble swarm is visualized using PIV in the measurement section. The size of the measurement section is 40 mm $\times$ 40 mm. A sample snapshot of a bubbly flow with gas volume fraction of $\alpha \approx 0.025$ in Newtonian fluid with viscosity (b) Re = 626 and We = 3.2, and (c) Re = 6 and We = 0.5, respectively. The scale bar is 10 mm. }
    \label{fig:ExperimentalSetup}
\end{figure}

Figure \ref{fig:ExperimentalSetup}(a) shows the schematic of the experimental setup used in the present study. It consists of a transparent acrylic tank of height 1000 mm with a cross-section of 100 mm $\times$ 50 mm. Monodispersed air bubbles are injected at the bottom of the tank through a removable capillary bank. The capillary bank is custom made with identical capillaries of inner diameter 0.6 mm in tandem with a secondary capillary of inner diameter 0.25 mm, arranged in a hexagonal array to increase the number of identical capillaries per unit area \citep{martinez2007measurement}. The tandem arrangement provides the sufficient hydraulic resistance through the capillaries such that individual bubbles are formed in a quasi-steady manner, thereby avoiding the generation of gas jets \citep{oguz1993dynamics}. The gas flow rate is adjusted using a needle valve.

The velocity fluctuations in the liquid phase is measured from the wake behind the bubble swarm, following the method used by \citet{riboux2010experimental}. The airflow is abruptly stopped by using a rapid solenoid valve. Then, the region behind the bubble swarm is studied using high-speed particle image velocimetry (Photron FASTCAM SA5 at 500 frames per second). The recorded images were then analyzed using PIVLab in MATLAB. For the PIV analysis, 32 $\times$ 32 pixels interrogation regions in the first pass and 16 $\times$ 16 pixels interrogation regions in the second pass with 50\% overlap on the subsequent pass is used. \textcolor{black}{The physical distance between two neighbouring vectors is 0.3 mm.} Spurious vectors are detected by median test and replaced by interpolating neighbor vectors. From the 2D PIV data, the liquid velocity fluctuations in the bubble swarm wake are obtained as follows $\mathbf{u'} = \mathbf{u} - \langle \mathbf{u}\rangle$. Here $\langle \rangle$ represent the average in space. The energy spectra, $E(k)$, of the velocity fluctuations are determined from the decaying turbulence behind the bubble swarm following \cite{riboux2010experimental}. The energy spectra of horizontal and vertical velocity fluctuations are then obtained using the Welch method by averaging the energy spectra of each rows and columns, respectively. To determine the velocity fields within the bubbly swarm at low gas volume fraction, fluorescent particles with an orange filter in the camera was used. The details of the fluorescent particle imaging technique can be found from our earlier study \citep{ravisankar2022hydrodynamic}.

\subsection{Test fluids}

To prepare viscous Newtonian fluids,  water-glycerin mixtures were used. The properties of the fluids used in the study, shear viscosity, density and surface tension were measured using ARES-G2 Rheometer (TA Instruments), Density meter (Anton Paar) and Bubble pressure tensiometer (KRUSS Scientific Instruments), are listed in Table \ref{table:FluidPropertiesLowRE}. To reduce bubble coalescence, a small amount of magnesium sulfate salt (0.05 mol/l) was added to all the fluids used in the study, following \citet{lessard1971bubble}. The relevant dimensionless numbers used in the current study are (i) Reynolds number, Re = $\rho U D/\mu$, where $U$ is the average bubble velocity, $D$ is the average bubble diameter, $\rho$ is the fluid density, and $\mu$ is the fluid viscosity, and (ii) Weber number, We = $\rho U^2 D/\sigma$, where $\sigma$ is the surface tension. The average bubble diameter and velocity were obtained from the probability distribution obtained from the image processing. The Kolmogorov's length scale was calculated as $\eta = (\nu^3/\epsilon)^{1/4}$, where $\nu = \mu/\rho$ is the kinematic viscosity and $\epsilon$ is the dissipation rate. In turn, $\epsilon$ determined from the fluctuating rate of strain tensor using local isotropic assumption  following \citet{xu2013accurate}. The mean gas volume fraction, $\alpha = (H_0/\Delta H + 1)^{-1}$, is measured from the increase in the liquid level after the injection of bubbles, where $H_0$ is the initial liquid level and $\Delta H$ is the liquid level increase.


\begin{table}
\centering
\begin{tabular} { p{2.3cm} p{1.2cm} p{1.2cm} p{1cm} p{1.4cm} p{1.2cm} p{1cm} p{0.8cm} p{0.8cm}}
Fluids & $\rho$ & $\sigma$ & $\mu$  & $D$ & $U$ & $\eta$ & Re & We \\ 
(in water)&  (kg/m$^{3}$) & (mN/m) & (Pa.s) & (mm)& (mm/s) & (mm)&& \\
\\
\tikzcircle[fill=color1]{4pt} 10\% Glycerin &1003.0 & 73.23 & 0.001  & 2.8 $\pm$ 0.2 & 290  & 0.10 & 626 & 3.2 \\
\tikzcircle[fill=color2]{4pt} 50\% Glycerin  &1140.6 & 72.38 & 0.005  & 2.6 $\pm$ 0.1 & 243  &0.43 & 144 & 2.1 \\
\tikzcircle[fill=color3]{4pt} 60\% Glycerin   & 1142.1 & 71.97 & 0.007  & 2.6 $\pm$ 0.1 & 206 &0.47 & 87 & 1.8 \\
\tikzcircle[fill=color4]{4pt} 75\% Glycerin  &1195.3 & 70.07 & 0.029  & 2.5 $\pm$ 0.2 & 114 &1.15 & 11 & 0.6 \\
\tikzcircle[fill=color5]{4pt} 85\% Glycerin   & 1224.0 & 70.04 & 0.048  & 2.3 $\pm$ 0.1 & 109  &1.45 & 6 & 0.5 \\
\end{tabular}
\caption[Table 2]{Physical properties of the fluids: $\rho$ - density; $\sigma$ - surface tension; $\mu$ - viscosity; \\$D$ - average bubble diameter; $U$ - average bubble velocity; $\eta$ - Kolmogorov's length scale;\\  Re - Reynolds number; We - Weber number.}
\label{table:FluidPropertiesLowRE}
\end{table}

\section{Results and Discussions}\label{sec:results}

\subsection{PDF of the velocity fluctuations}

\begin{figure}[h]
\centering
\subfigure[]{\mbox{\includegraphics[width=0.485\textwidth]{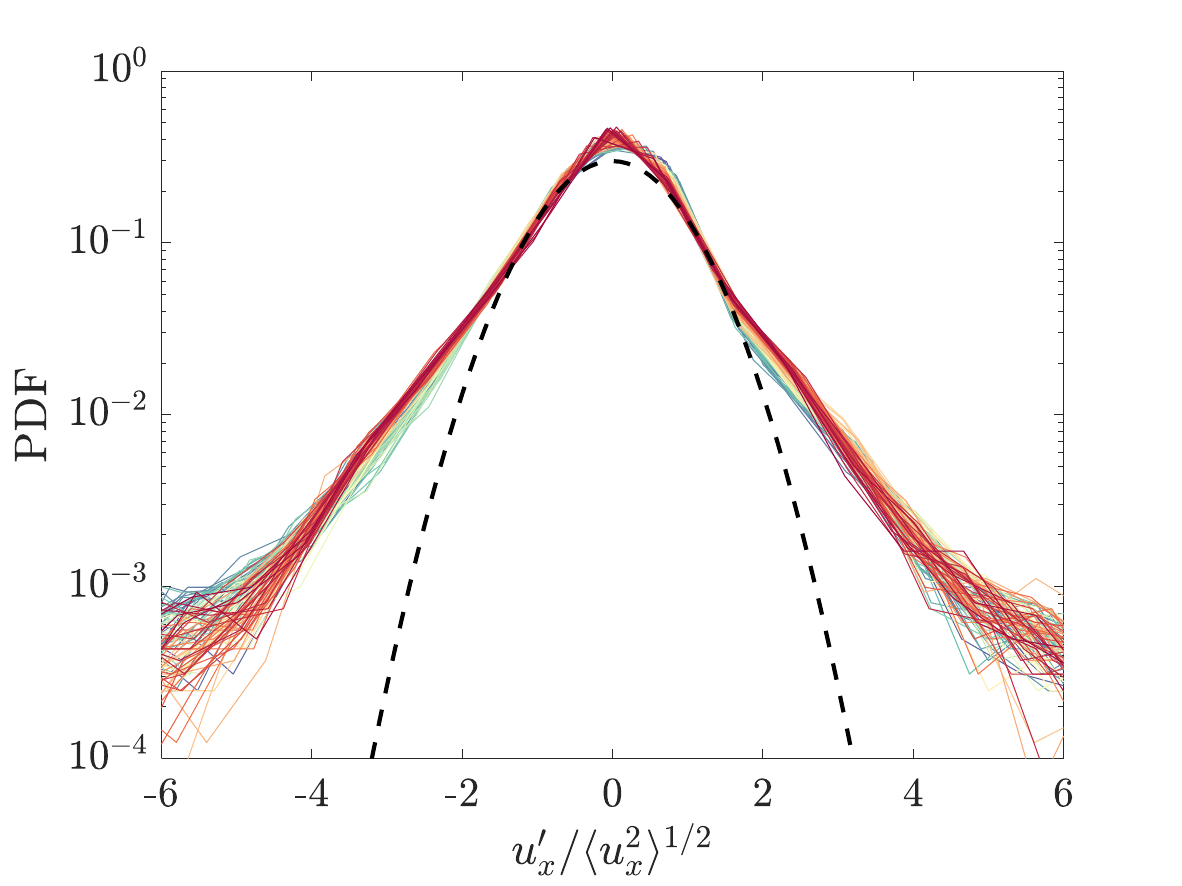}}} \,\,
\subfigure[]{\mbox{\includegraphics[width=0.485\textwidth]{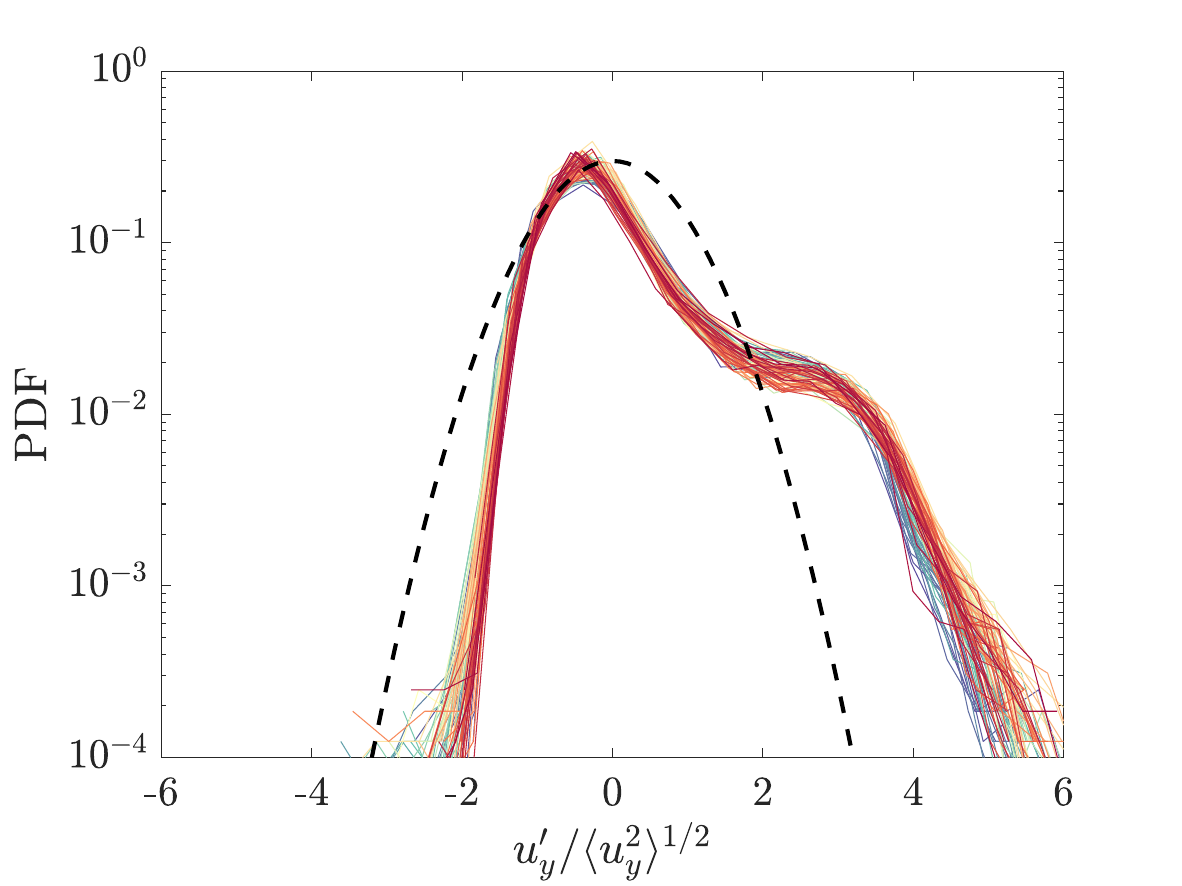}}}
\caption{PDF of the liquid velocity fluctuations (a) in the horizontal and (b) vertical direction within the bubbly flows in viscous Newtonian fluid (Re = 6) normalized by the standard deviation for gas volume fraction of $\alpha$ = 0.010. The dashed line corresponds to the Gaussian profile. \textcolor{black}{Here, the color gradients, ranging from blue to red, correspond to 2 ms of measurement within the bubble swarm}.}
\label{fig:PDF}
\end{figure}

Figure \ref{fig:PDF} shows the probability density functions (PDFs) of the horizontal and vertical velocity fluctuations within the bubble swarm normalized by the corresponding standard deviations for a bubble Reynolds number, Re = 6 at a constant gas volume fraction of $\alpha \approx 0.010$. The PDFs of the vertical velocity fluctuations are positively skewed whereas the PDFs of the horizontal velocity fluctuations are symmetric and non-Gaussian, similar to that of the bubbly flows at high Reynolds number \citep{risso2002velocity,almeras2017experimental,riboux2010experimental}. Here, determining the PDFs within the bubble swarm is essential as the velocity fields in the wake of the bubble swarm will be symmetric and Gaussian \citep{lee2021scale,ma2022experimental}. Though there is little to no wake for low Reynolds number bubbles, here, the exponential tail and the positive skewness in the vertical velocity fluctuations are attributed to the large fluctuations in the vicinity of the bubbles \citep{riboux2010experimental,almeras2017experimental}.

\subsection{Energy spectra of the velocity fluctuations}

\begin{figure}[h]
\centering
\subfigure[]{\mbox{\includegraphics[width=0.55\textwidth]{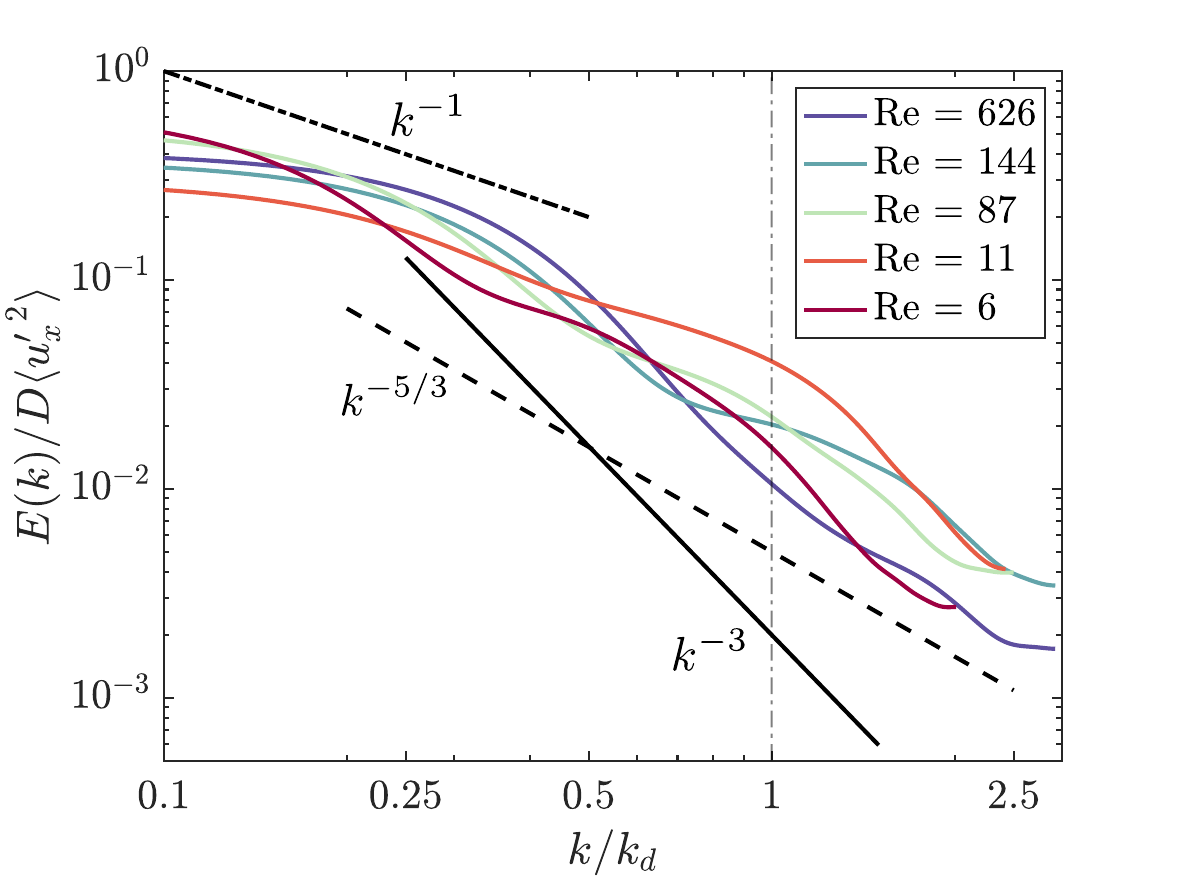}}} \,\,
\subfigure[]{\mbox{\includegraphics[width=0.55\textwidth]{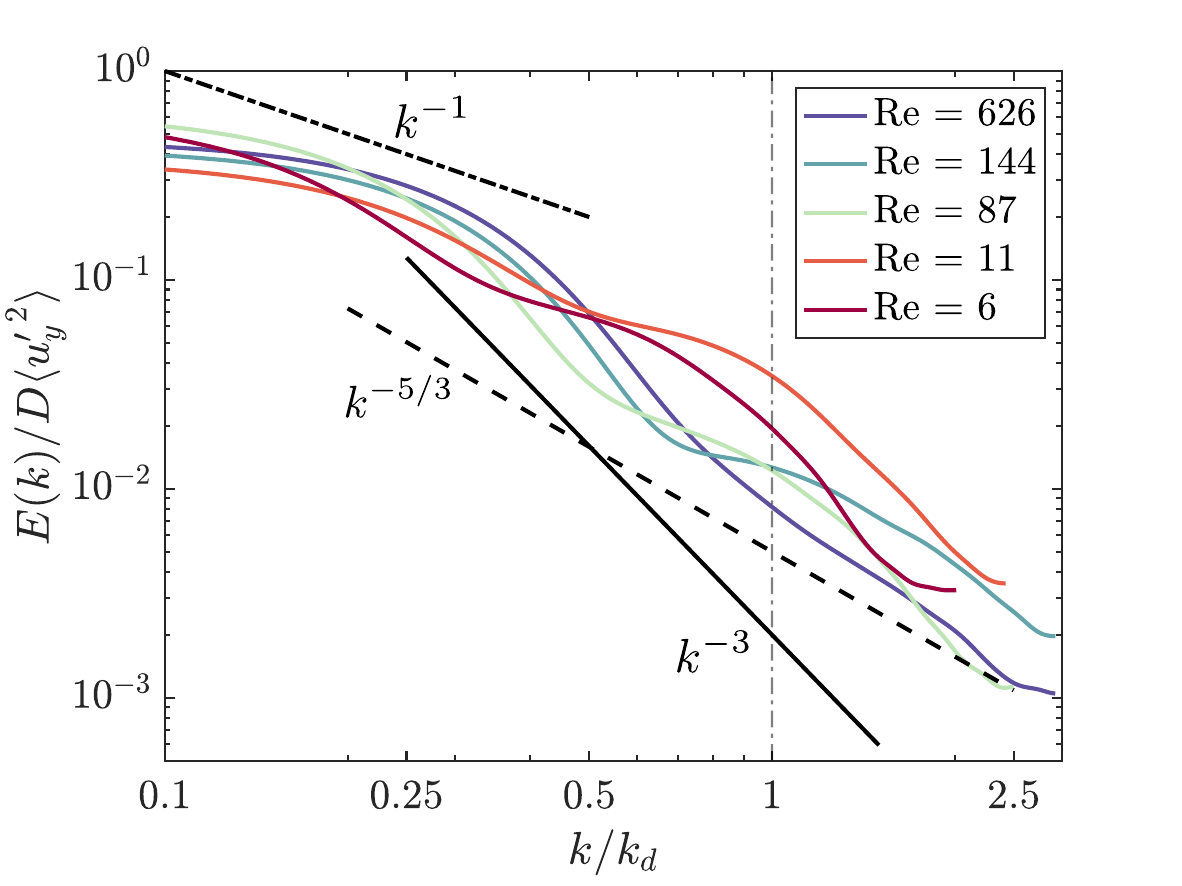}}}
\caption{(a) Horizontal, and (b) vertical spectra of the liquid velocity fluctuations normalized by the bubble diameter and the variances for a family of Reynolds number at a constant gas volume fraction of $\alpha \approx 0.025$. The abscissa is normalized by the wavenumber corresponding to the bubble diameter, $k_d$. The solid, dashed and dot-dashed black lines correspond to the $k^{-3}$, $k^{-5/3}$, and $k^{-1}$ scaling, respectively.}
\label{fig:Spectra}
\end{figure}

We investigate the energy spectra of velocity fluctuations for different bubble Reynolds number at a fixed gas volume fraction. Figure \ref{fig:Spectra} shows the horizontal and vertical energy spectra of velocity fluctuations normalized by the corresponding variance and bubble diameter for a family of bubble Reynolds number at a constant gas volume fraction of $\alpha \approx 0.025$. Here, the abscissa is normalized by the wavenumber corresponding to the bubble diameter, $k_d = 2 \pi/D$. As seen in the Fig. \ref{fig:Spectra}, at large scales (small wavenumbers) the energy spectra of the liquid velocity fluctuations scales as $k^{-1}$ in agreement with \citet{zamansky2024turbulence}. At the intermediate scales, for Reynolds numbers, Re $\gg$ $O$(10), the signature $k^{-3}$ scaling of the energy spectra is obtained similar to \citep{martinez2007measurement,riboux2010experimental}, which also serves as a validation of our experimental method (see Appendix \ref{appB}). When the bubble Reynolds number Re $\sim O(100)$, the $k^{-3}$ subrange is significantly narrower compared to that when Re = 626. On further decreasing the Reynolds number, for Re = 11 and 6, the signature $k^{-3}$ scaling for the pseudoturbulence does not emerge. The numerical results obtained for the bubble induced turbulence by \citet{bunner2002dynamics2} for Re between 12 and 30 found energy spectra decays with a $k^{-3}$ scaling, which agrees with the present findings. To our knowledge no other data for Re $<$ $O$(10) exists.

It has been argued that the observed $k^{-3}$ scaling is the result of the contribution from the bubble wakes \citep{risso2018agitation}. In the current study, however, for $O(1) < $ Re $ < O(10)$, no significant wake behind the bubbles are expected to appear \citep{blanco1995structure}. Thus, instead of the signature $k^{-3}$ scaling for the pseudoturbulence, the energy spectra of velocity fluctuations induced by the low Reynolds number bubbles in our experiments scales as $k^{-5/3}$. We argue that $k^{-5/3}$ scaling emerges from the small scale disturbances generated by the bubble motion, which are strong enough to generate a large scale flow with the same characteristics as the Kolmogorov's turbulence \citep{mazzitelli2009evolution}. Thus, we provide the first experimental evidence that at very low Reynolds numbers bubbly flows, the energy spectra of the liquid velocity fluctuations scales as $k^{-5/3}$. Further, we note that at small scales (large wavenumbers) the energy spectra of the liquid velocity fluctuations for $O(1) < $ Re $ < O(10)$ \textcolor{black}{follows an exponential decay} due to the dominance of viscosity beyond the Kolmogorov's microscale, $\eta$, as listed in the Table \ref{table:FluidPropertiesLowRE}. The compensated energy spectra, included in Appendix \ref{appA},  show the emergence of $k^{-3}$ and $k^{-5/3}$  more clearly at different length scales.


\begin{figure}[h]
\centering
\subfigure[]{\mbox{\includegraphics[width=0.485\textwidth]{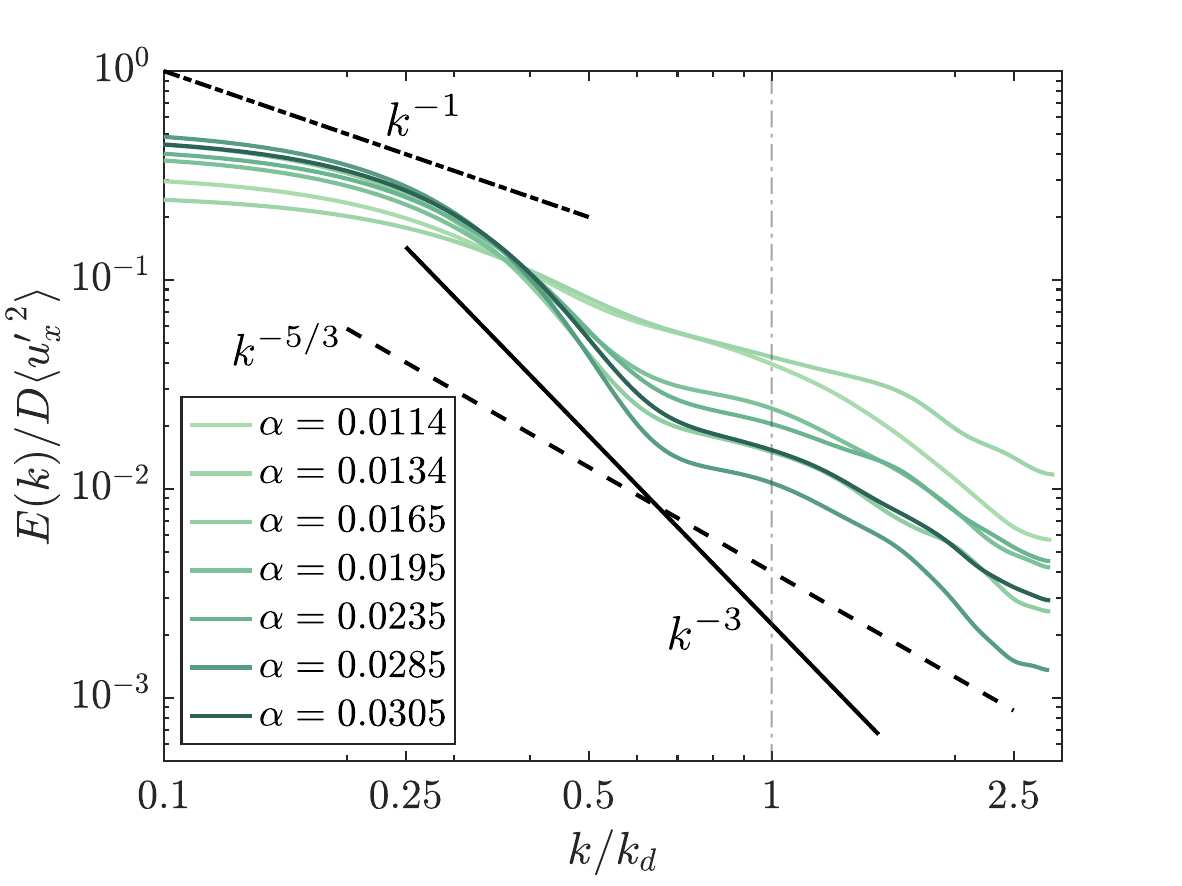}}} \,\,
\subfigure[]{\mbox{\includegraphics[width=0.485\textwidth]{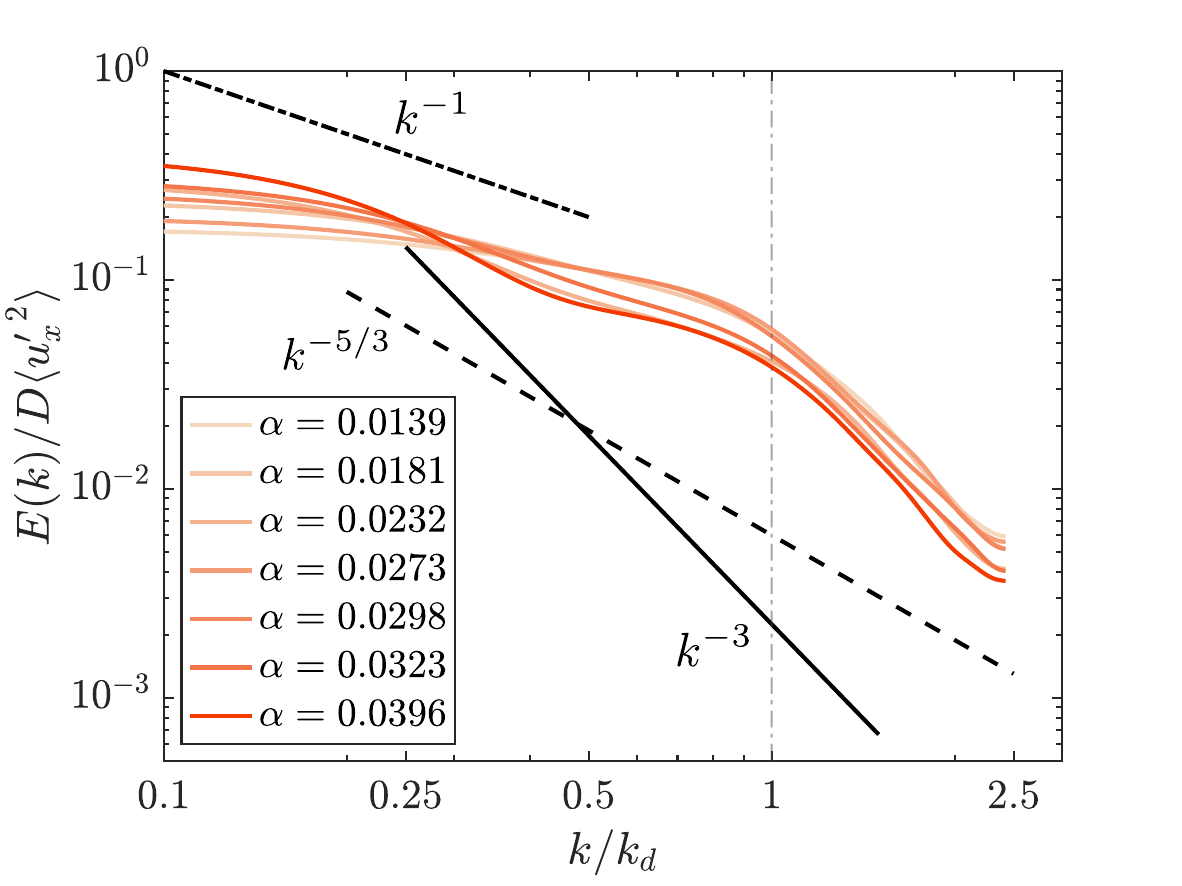}}} 
\caption{\justifying Horizontal spectra of the liquid velocity fluctuations normalized by the bubble diameter and the variances for a family of gas volume fraction at (a) Re = 114 and (b) Re = 11. The abscissa is normalized by the wavenumber corresponding to the bubble diameter, $k_d$. The solid, dashed and dot-dashed black lines correspond to the $k^{-3}$, $k^{-5/3}$, and $k^{-1}$ scaling, respectively.}
\label{fig:GasVolumeSpectra}
\end{figure}

Further, \citet{riboux2010experimental} suggested that the $k^{-3}$ scaling observed in the pseudoturbulence does not depend on the bubble diameter and gas volume fraction for a wide range of Reynolds numbers. We show that this is not the case at small Reynolds number. Figure \ref{fig:GasVolumeSpectra}(a) shows the horizontal spectra of liquid velocity fluctuations normalized by the bubble diameter and the variance for a family of gas volume fraction at Re = 114. It is to be noted that at very low gas volume fractions, though the bubbles have significant wake, the energy spectra does not show the $k^{-3}$ scaling. However, as the gas volume fraction increases, the $k^{-3}$ scaling emerges. Note that this $k^{-3}$ subrange is significantly narrower. This is because as the number of bubbles (i.e the gas volume fraction) increases the average inter distance between the bubbles decreases, thus leading to wake interactions. This is further corroborated using spatial correlation in the next section. Whereas, for Re = 11 as seen in the Fig. \ref{fig:GasVolumeSpectra}(b), the $k^{-3}$ scaling is not observed, instead the $k^{-5/3}$ scaling is obtained. This is in agreement with the results from \citet{mazzitelli2003effect} that the slope of the energy spectra no longer show the signature $k^{-3}$ scaling for the pseudoturbulence and that the slope of the energy spectra depend on the number of bubbles and therefore is non-universal. 

\subsection{\textcolor{black}{Structure functions}}

\begin{figure}[h]
    \centering
    \includegraphics[scale=0.4]{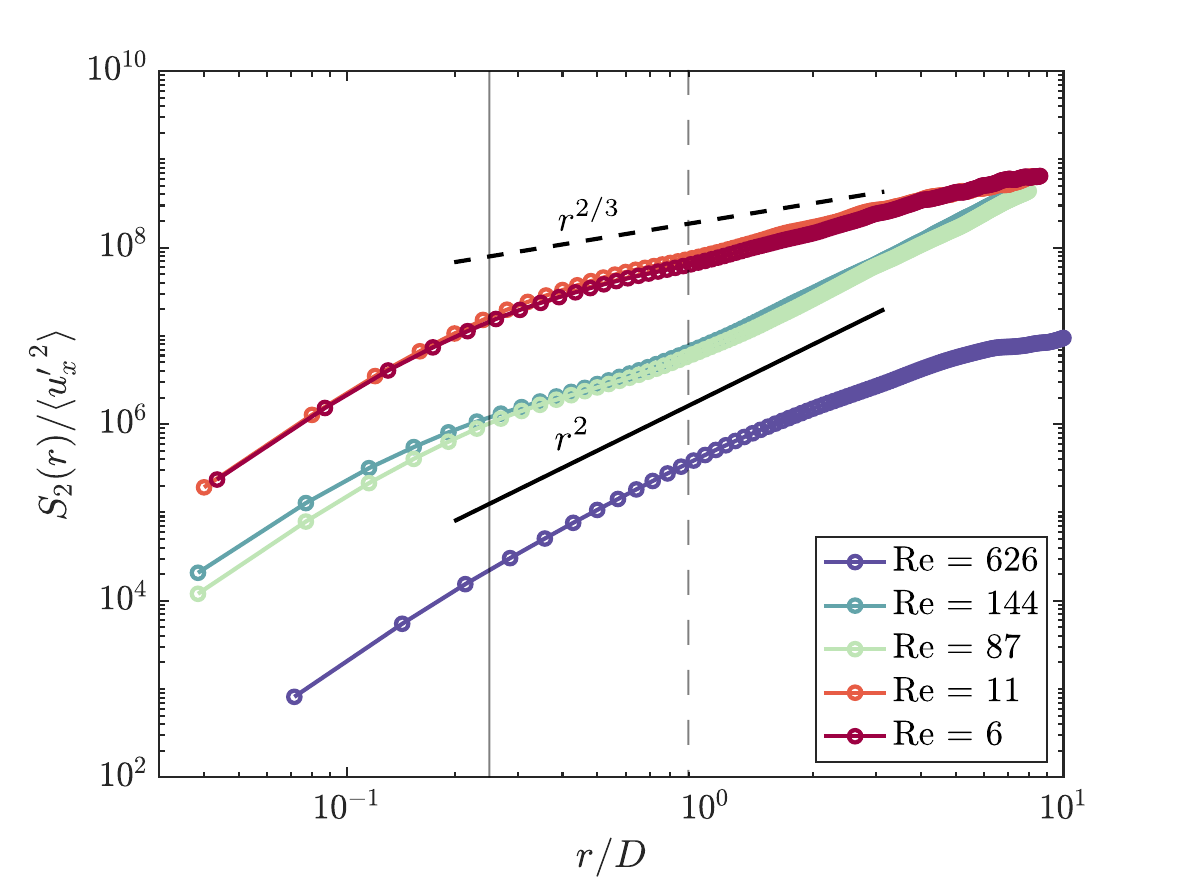}
    \caption{\justifying \textcolor{black}{The normalized second-order structure function, $S_2(r)$ as a function of normalized length scale $r$ for a family of Reynolds numbers at a constant gas volume fraction of $\alpha \approx 0.025$. The solid and dashed lines correspond to $r^{2}$ and $r^{2/3}$ scaling, respectively.}}
    \label{fig:StructureFunction}
\end{figure}

\textcolor{black}{To demonstrate that at low Reynolds numbers, the energy spectra of the bubbly flows exhibit a $k^{-5/3}$ scaling, we consider the second-order longitudinal velocity structure function across a length scale $r$ defined as,}
\begin{equation}
\color{black}
    S_2(r) = \langle { [\mathbf{u_x(x+r) - u_x(x)}].(\mathbf{r}/r)}^2 \rangle.
\end{equation} \textcolor{black}{Here the $\langle \rangle$ denotes the spatial average. For a homogeneous and isotropic turbulence in a Newtonian incompressible fluids, it is known that if the second-order velocity structure function scales as $S_2(r) \sim r^\beta$, then the corresponding energy spectra should scale as $E(k) \sim k^{-(\beta+1)}$ \citep{frisch1995turbulence}. For classical Kolmogorov turbulence, the second-order velocity structure function scales as $S_2(r) \sim r^{2/3}$ in the inertial range and as $S_2(r) \sim r^{2}$ in the dissipative range. Figure \ref{fig:StructureFunction} shows the normalized second-order velocity structure function observed in the wake of the bubble swarm as a function of normalized length scale $r$, across a range of Reynolds numbers for a constant gas volume fraction of $\alpha \approx 0.025$. For all experiments, when the normalized length scale $r/D < 0.25$ (dissipative range), the structure function scales as $S_2(r) \sim r^{2}$. In the inertial range, because of the bubble wake-wake interactions, instead of the $S_2(r) \sim r^{2/3}$ scaling, the second-order structure function scales as $S_2(r) \sim r^2$ for Re $\gg O(100)$ \citep{ma2022experimental}. At Reynolds numbers around Re $\sim O(100)$, for length scales smaller than the bubble diameter ($r/D < 1$), the structure function scales as $S_2(r) \sim r^{2/3}$. Whereas, for length scales larger than the bubble diameter ($r/D > 1$), the structure function scales as $S_2(r) \sim r^2$. This highlights that the $k^{-3}$ scaling in the energy spectra can be recovered for bubbles with sufficiently large wakes (i.e. Re = 144 and Re = 87). On further decreasing the Reynolds number Re $\sim$ O(10), the structure function follows $S_2(r) \sim r^{2/3}$ for length scales larger than that of the dissipative range and thus the corresponding energy spectra of the velocity fluctuations at low Reynolds number bubbly flows scales as $E(k) \sim k^{-5/3}$, as shown in Fig. \ref{fig:Spectra}.}

\subsection{Spatial correlation}

\begin{figure}[h]
\centering\subfigure[]{\mbox{\includegraphics[width=0.75\textwidth]{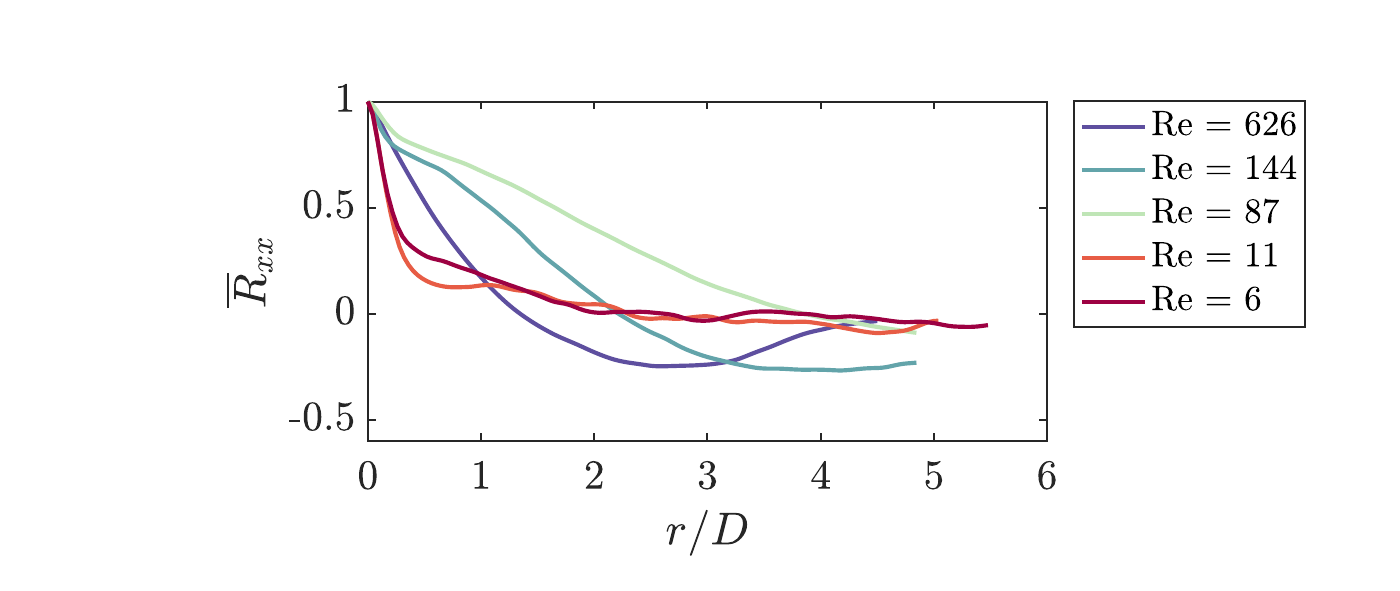}}} \,\,
\subfigure[]{\mbox{\includegraphics[width=0.485\textwidth]{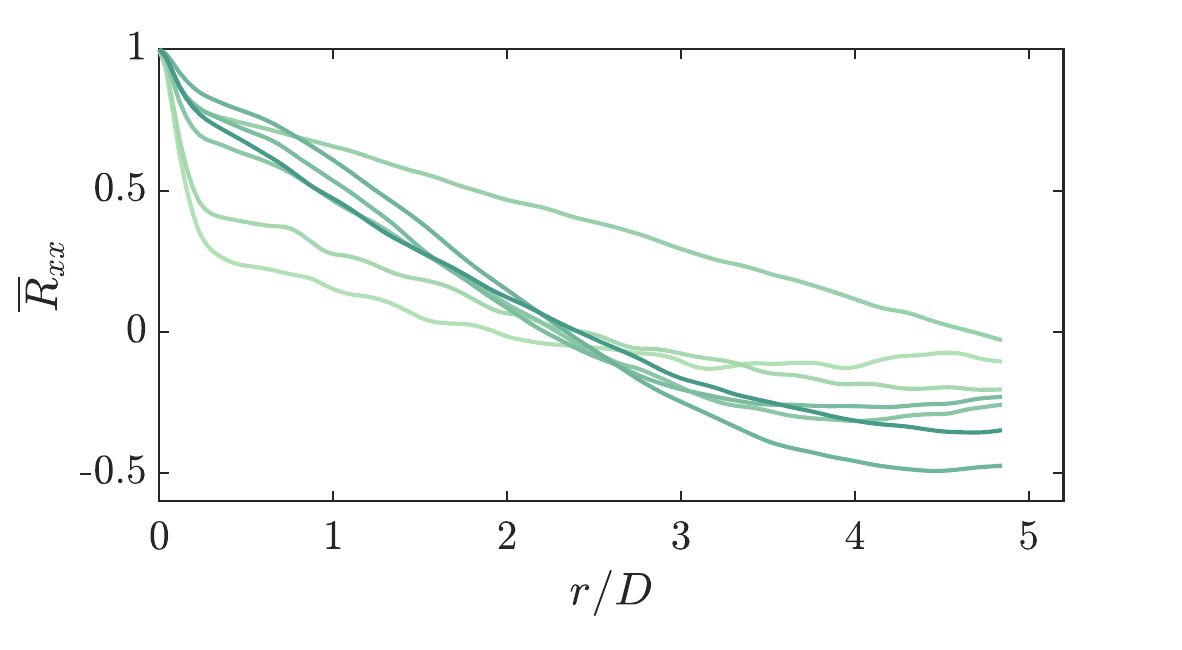}}} \,\,
\subfigure[]{\mbox{\includegraphics[width=0.485\textwidth]{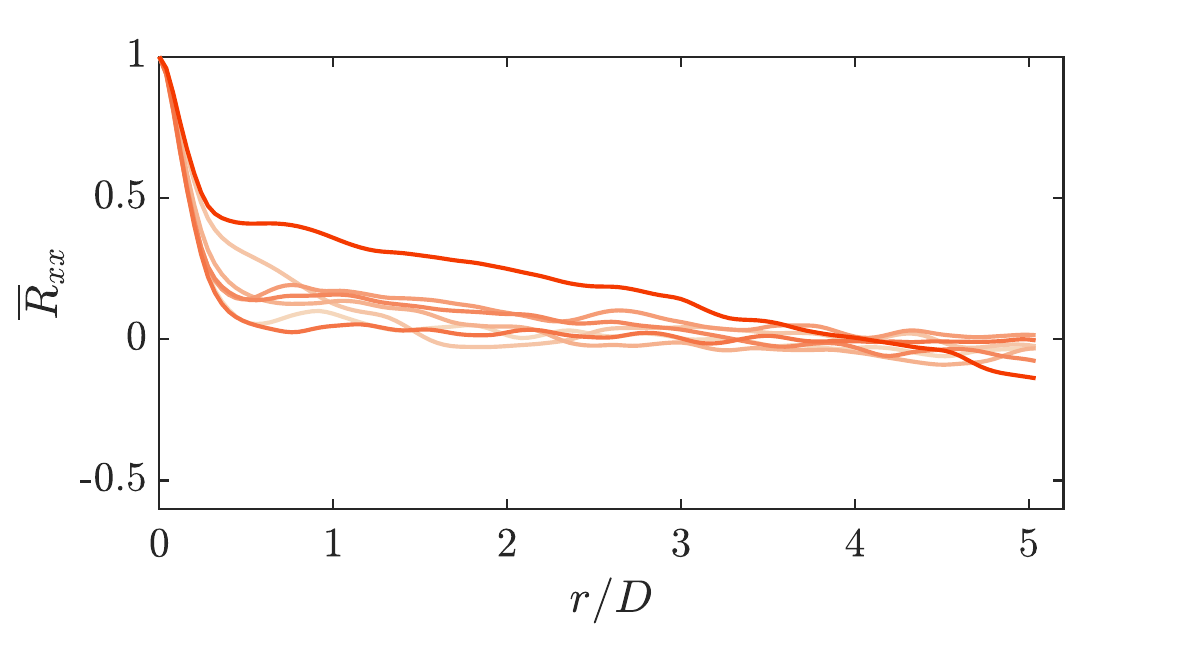}}}
\caption{Spatial correlation of the horizontal velocity observed in the wake of the bubble swarm (a) at a constant gas volume fraction of $\alpha \approx 0.025$ for a family of Reynolds number. Spatial correlation of the horizontal velocity for a family of gas volume fractions (mentioned in Fig. \ref{fig:GasVolumeSpectra}) at (b) Re = 114 and (c) Re = 11.}
\label{fig:GasVolumeCorrelation}
\end{figure}

To further understand why the slope of the energy spectra depends on the gas volume fraction for low to moderate Reynolds number, the spatial correlation of the horizontal velocity fields are considered.  The spatial correlation of the horizontal velocity, $\mathbf{u}_x$ defined as,

\begin{equation}
    {\overline{R}}_{xx} = \frac{\langle \mathbf{u}_x(\mathbf{x}) \cdot \mathbf{u}_x(\mathbf{x + r})\rangle}{\langle u^2_x \rangle}.
\end{equation} Here $\langle \rangle$ represent the average in space. Figure \ref{fig:GasVolumeCorrelation}(a) shows the spatial correlation of the horizontal velocity behind the bubble swarm at a constant gas volume fraction of $\alpha \approx 0.025$. It is evident that as the Re decreases from 626 to 144 to 87, we observe a higher correlation over longer distances. This may be because the size of the bubble decreases as fluid viscosity increases (see Table \ref{table:FluidPropertiesLowRE}). Therefore, more number of bubbles are needed to reach a fixed gas volume fraction (say $\alpha \approx 0.025$). This longer correlation with the decrease in the Reynolds number is in agreement with the numerical results from \citet{esmaeeli1996inverse}, who showed that there is an emergence of flow structures many times larger than the bubble size, which is also observed in particle suspensions at finite Reynolds numbers \citep{climent2003numerical}. However, on further decreasing the Reynolds number (Re = 11 and 6), the velocity is decorrleated quickly as seen in the Fig. \ref{fig:GasVolumeCorrelation}(a). This agrees with the previous literature \citep{lance1991turbulence,riboux2010experimental,risso2018agitation,mazzitelli2003effect} that the presence of wake is essential for the signature $k^{-3}$ scaling to emerge as seen in the Fig. \ref{fig:Spectra}. Figure \ref{fig:GasVolumeCorrelation}(b) shows the spatial correlation of the horizontal velocity observed behind the bubble swarm for a family of gas volume fractions at Re = 114. It is immediately evident that as the gas volume fraction increases, the horizontal velocity is correlated over a longer distance. Thus, agreeing with our results for the energy spectra, as seen in Fig. \ref{fig:GasVolumeSpectra}(a), dependence on the number of bubbles. However, as seen in the Fig. \ref{fig:GasVolumeCorrelation}(c) when the Re = 11, even with higher gas volume fraction, the horizontal velocity is quickly decorrelated beyond one bubble radius. 

\section{Conclusions} \label{sec:conclusions}

The bubbly flow properties in viscous Newtonian fluids have been studied experimentally by varying the concentration of glycerin in water mixture. Particle image velocimetry (PIV) technique was used to visualize the wake behind the bubble swarm to determine the velocity fluctuations in the decaying agitations. We demonstrated experimentally that the signature $k^{-3}$ scaling of the pseudoturbulence is replaced by $k^{-5/3}$ scaling for the energy spectra of velocity fluctuations induced by low Reynolds number bubbles. We showed that as the Reynolds number decreases to Re $\sim O(100)$, the $k^{-3}$ subrange becomes significantly narrower. Further, for low Reynolds bubbly flows the slope of the energy spectra depends on the number of bubbles in the flow. 

These experimental results agree with the numerical results by \citet{mazzitelli2003effect} that the energy spectra is non-universal for bubbles with $O(1) < $  Re  $ < O(10)$. To understand why the slope of the energy spectra depends on the gas volume fraction, the spatial correlation of the velocity field was considered. For a constant gas volume fraction, as the $O(10) < $  Re  $< O(100)$, a higher correlation over longer distances was observed in agreement with the results from \citet{esmaeeli1996inverse}. As the bubble size decreases with Reynolds number, more number of bubbles (i.e. higher gas volume fraction) are required to maintain the same gas volume fraction. Thus, as the inter-bubble distance among bubbles decreases the wake interactions are more pronounced. Whereas for $O(1) < $  Re  $ < O(10)$ the velocity quickly decorrelated beyond the bubble radius, thus the signature $k^{-3}$ scaling for the bubble induced turbulence does not emerge. 

Declaration of Interests. The authors report no conflict of interest.

\appendix

\section{Comparison with the experiments of \citet{riboux2010experimental}}\label{appB}

\begin{figure}[h]
    \centering
    \includegraphics[scale=0.4]{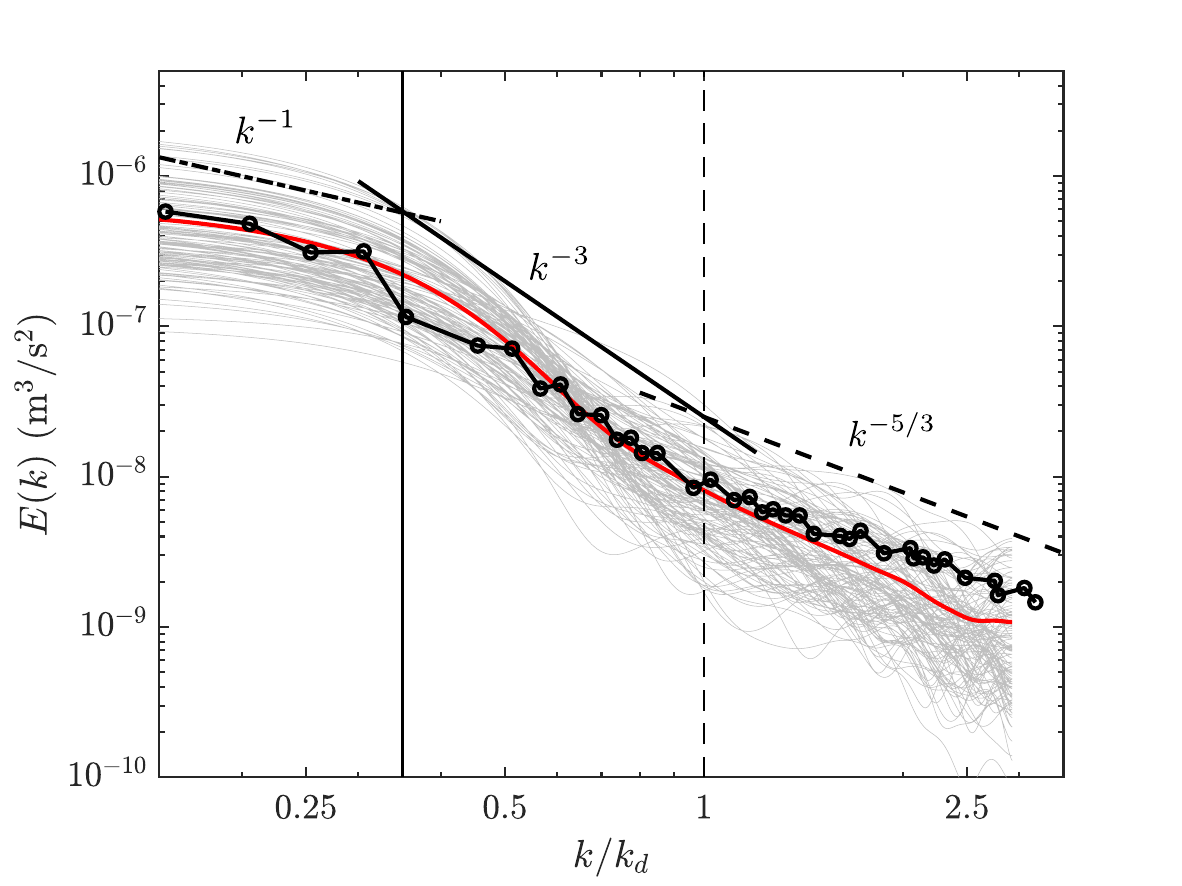}
    \caption{\justifying Vertical spectra for each vertical column of the liquid velocity fluctuations at Re = 626 and gas volume fraction, $\alpha \approx 0.0025$ is shown by the gray lines. The solid red line corresponds to the average energy spectra of the vertical velocity fluctuations. The abscissa is normalized by the wavenumber corresponding to the bubble diameter, $k_d$. Solid black line with symbols correspond to the spectrum obtained from experiments by \citet{riboux2010experimental} at Re = 670 and gas volume fraction, $\alpha \approx 0.0046$.}
    \label{fig:averagePSD}
\end{figure}

In Fig. \ref{fig:averagePSD}, we compare our measurement technique with that by \citet{riboux2010experimental}. The gray lines correspond to the energy spectra of the vertical liquid velocity fluctuations for each column of the velocity field obtained from the PIV measurement. The average of all these vertical energy spectra is shown as a solid red line. Here, we compare our result with that obtained by \citet{riboux2010experimental} at Re = 670 and gas volume fraction, $\alpha \approx 0.046$ (solid black line with symbols). Further, we clearly identify the $k^{-1}$ scaling at large scales (dot-dashed black line), $k^{-3}$ scaling in the intermediate scales (solid black line) and $k^{-5/3}$ scaling in the small scales (dashed black line) in accordance to \cite{zamansky2024turbulence}.

\section{Compensated energy spectra of velocity fluctuations}\label{appA}

\begin{figure}[h]
\centering
\subfigure[]{\mbox{\includegraphics[width=0.485\textwidth]{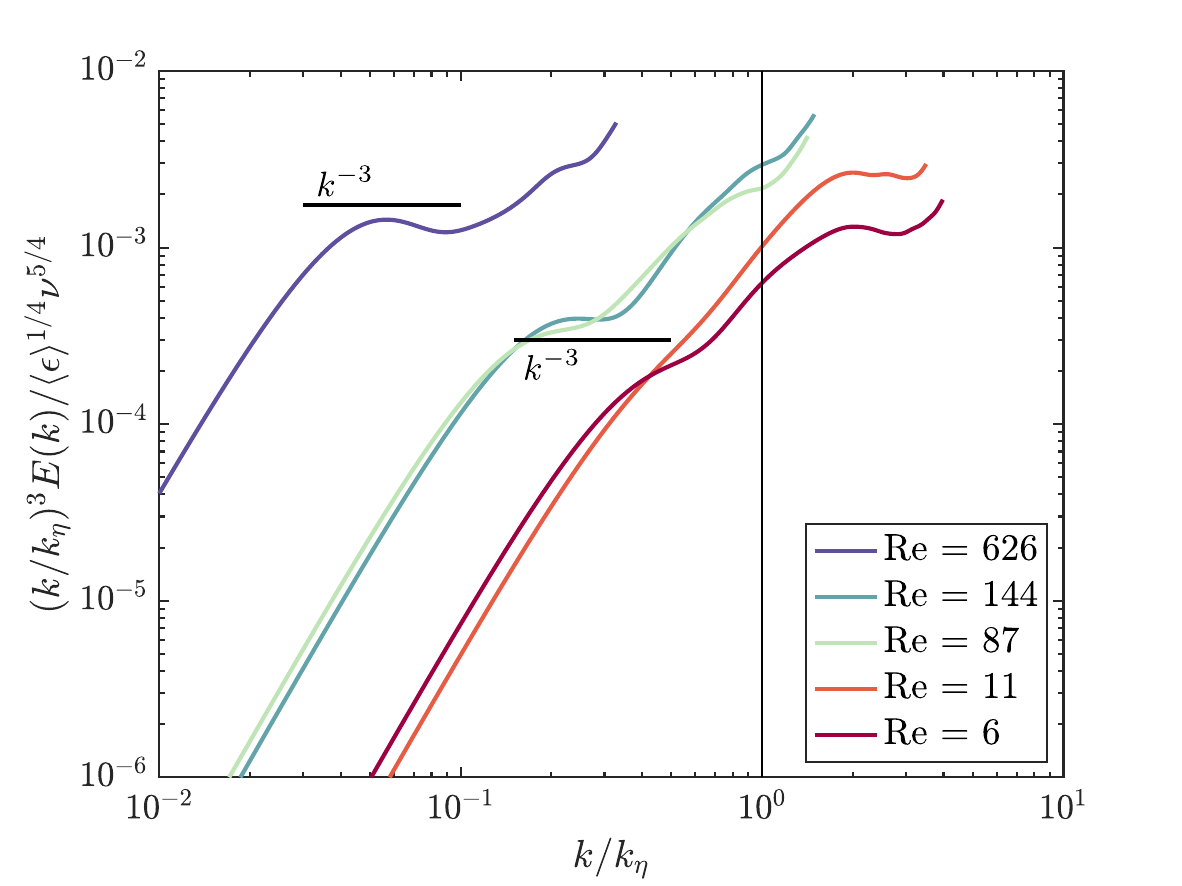}}} \,\,
\subfigure[]{\mbox{\includegraphics[width=0.485\textwidth]{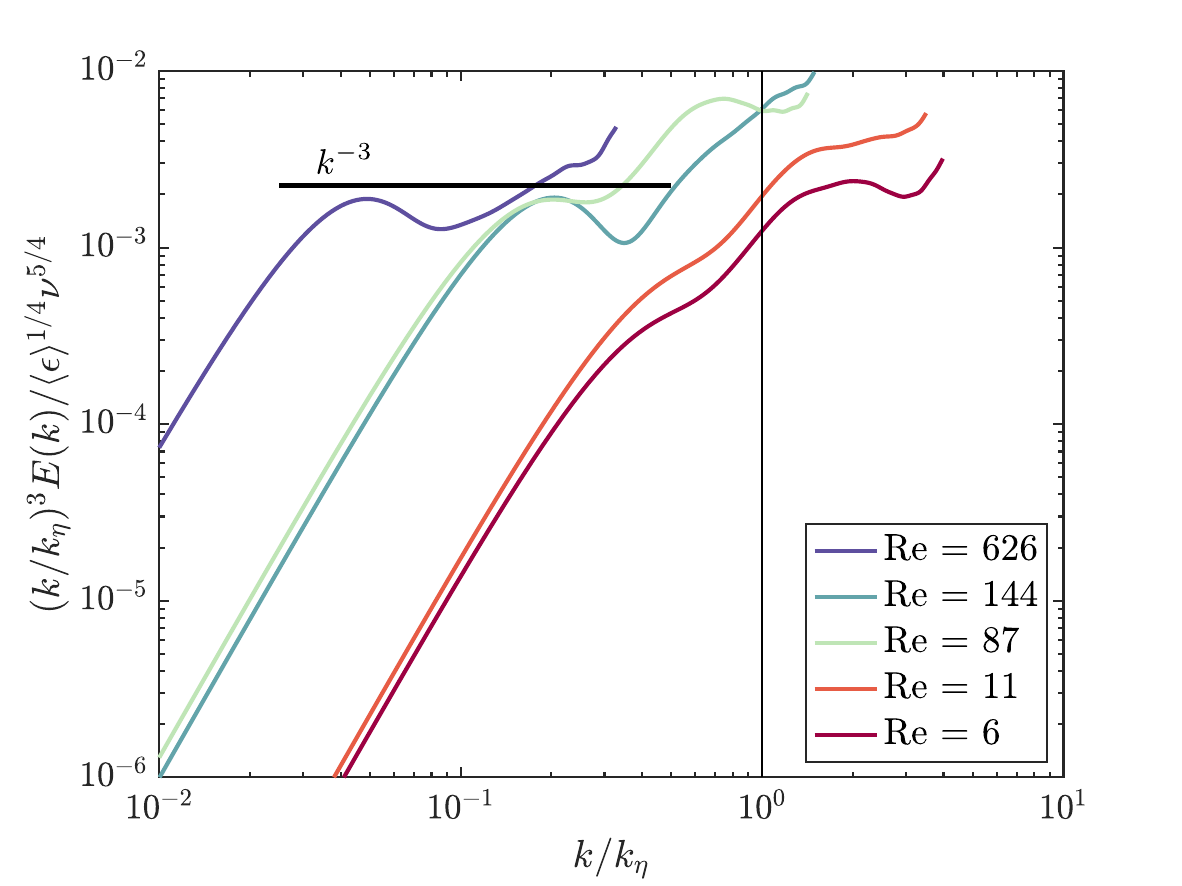}}}
\caption{\textcolor{black}{Compensated energy spectra of the (a) horizontal and (b) vertical liquid velocity fluctuations showing the emergence of $k^{-3}$ scaling for a family of Reynolds number at a constant gas volume fraction of $\alpha \approx $ 0.025. The solid line denote the $k^{-3}$ scaling.}}
\label{fig:ReplotSpectra3}
\end{figure}

\begin{figure}[h]
\centering
\subfigure[]{\mbox{\includegraphics[width=0.485\textwidth]{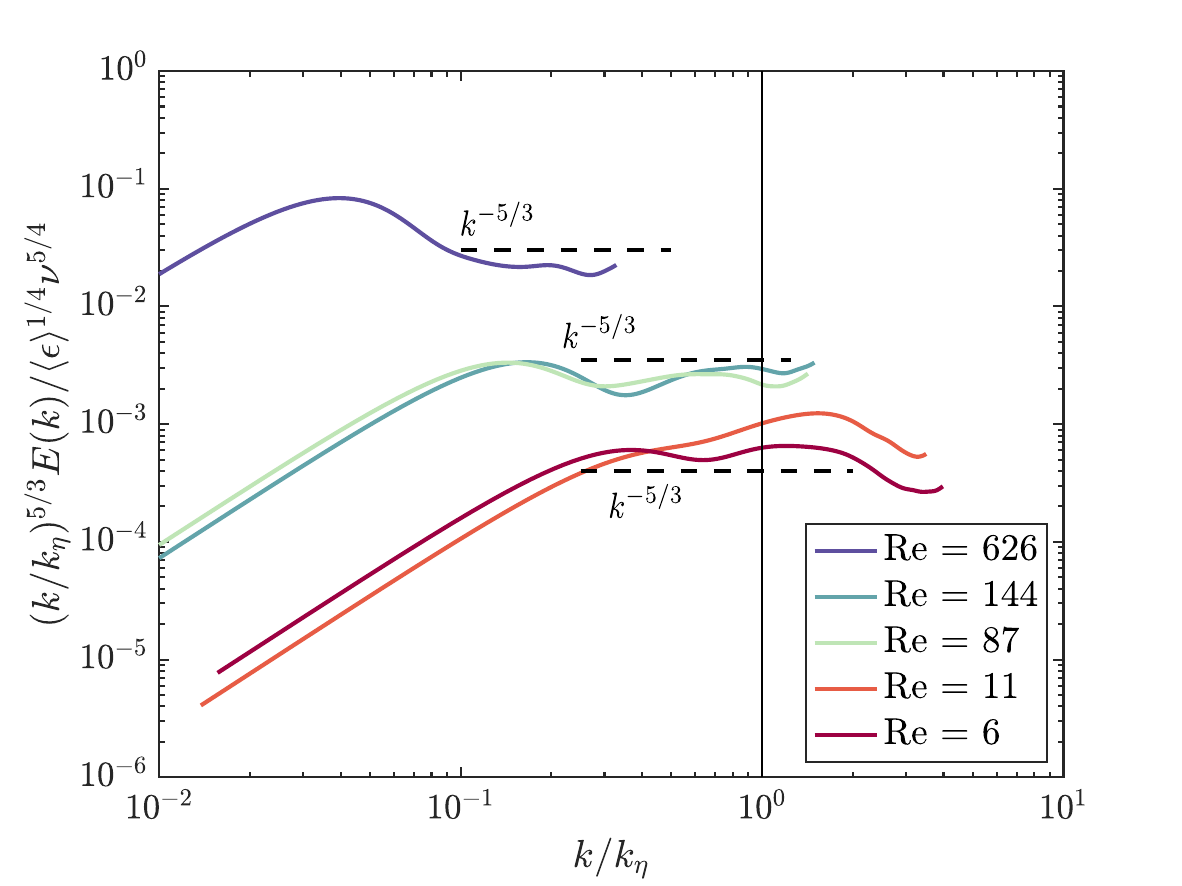}}} \,\,
\subfigure[]{\mbox{\includegraphics[width=0.485\textwidth]{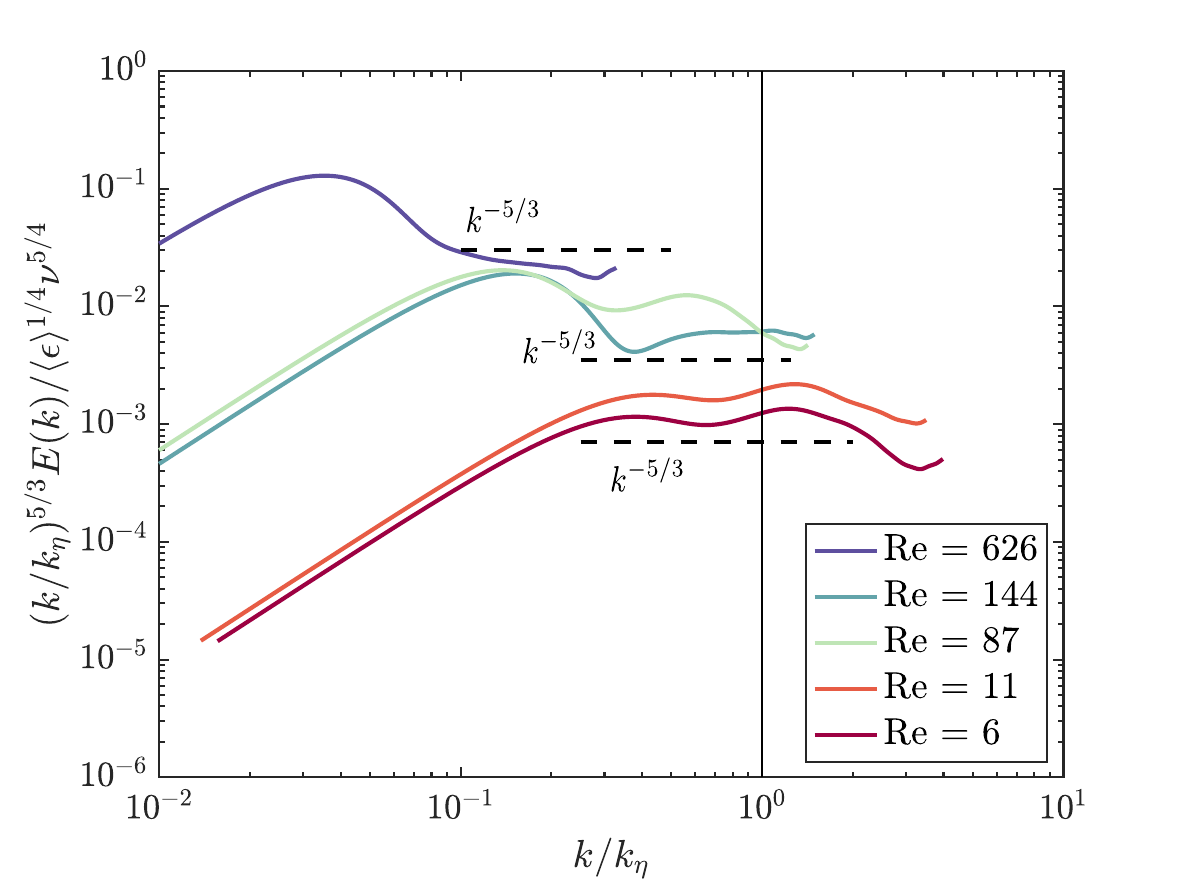}}}
\caption{\textcolor{black}{Compensated energy spectra of the (a) horizontal and (b) vertical liquid velocity fluctuations showing the emergence of $k^{-5/3}$ scaling for a family of Reynolds number at a constant gas volume fraction of $\alpha \approx $ 0.025. The dashed line denote the $k^{-5/3}$ scaling.}}
\label{fig:ReplotSpectra5-3}
\end{figure}

\textcolor{black}{In Fig. \ref{fig:ReplotSpectra3}, we plot the compensated spectra of the spectra of the horizontal and vertical liquid velocity fluctuations showing the emergence of the $k^{-3}$ scaling for a range of Reynolds number at a constant gas volume fraction of $\alpha \approx 0.025$. Here, the abscissa is normalized by the wavenumber corresponding to the the Kolmogorov length scale, $k_\eta = 2 \pi/\eta$. It is evident that the compensated energy spectra show the emergence of $k^{-3}$ scaling when the Re $\gg O(10)$. Whereas, when the Re $\sim O(10)$, the energy spectra of the liquid velocity fluctuations shows a $k^{-5/3}$ scaling as seen from the Fig. \ref{fig:ReplotSpectra5-3}. Note that for Re $\gg O(10)$, the $k^{-5/3}$ scaling is recovered for wavenumbers greater than that of the bubble diameter. }

\newpage
\bibliographystyle{jfm}
\bibliography{biblio}

\end{document}